\documentclass[11pt,psamsfonts]{amsart}
\usepackage[foot]{amsaddr}
\usepackage{amsmath,amsthm,amsfonts,amssymb}
\usepackage{eucal}
\usepackage{graphicx}
\usepackage{caption}
\usepackage{indentfirst}
\usepackage{bbm}
\usepackage{grffile}
\usepackage[all,knot]{xy}
\usepackage{chngcntr}
\usepackage{floatrow}
\usepackage{subfig}
\usepackage[colorlinks, citecolor=blue, linkcolor=blue]{hyperref}
\usepackage[usenames, dvipsnames]{color}
\usepackage{enumitem,kantlipsum}
\usepackage{amsbsy}
\usepackage{multicol}
\usepackage{multirow}
\usepackage{mathrsfs}
\usepackage[margin=1.3in]{geometry}
\counterwithin{figure}{section}
\xyoption{arc}

\newcommand{\A}{\mathcal{A}}
\newcommand{\M}{\mathcal{M}}

\DeclareMathOperator{\Hom}{Hom}

\DeclareMathOperator{\Fun}{Fun}

\newcommand{\C}{\mathbb C}
\newcommand{\mC}{\mathcal{C}}

\newcommand{\mZ}{\mathcal{Z}}
\newcommand{\mQ}{\mathcal{Q}}

\newcommand{\mfD}{\mathfrak{D}}
\newcommand{\Z}{\mathbb Z}

\newcommand{\B}{\mathcal{B}}

\newcommand{\comments}[1]{}

\newcommand{\ket}[1]{|#1\rangle}

\renewcommand{\Z}{\mathbb{Z}}

\newcommand{\overbar}[1]{\mkern 2.3mu\overline{\mkern-2.3mu#1\mkern-2.3mu}\mkern 2.3mu}

\numberwithin{equation}{section}

\theoremstyle{definition}

\begin{document}

\title{Topological Quantum Computation with Gapped Boundaries and Boundary Defects}
\author{Iris Cong$^{1,3}$}
\email{$^1$cong@g.harvard.edu}
\address{$^1$Dept of Physics, Harvard University\\
Cambridge, MA 02138\\
U.S.A.}

\author{Zhenghan Wang$^{2,3}$}
\email{$^3$zhenghwa@microsoft.com}
\address{$^2$Dept of Mathematics\\
    University of California\\
    Santa Barbara, CA 93106-6105\\
    U.S.A.}
\address{$^3$Microsoft Station Q\\
    University of California\\
    Santa Barbara, CA 93106-6105\\
    U.S.A.}

\begin{abstract}
We survey some recent work on topological quantum computation with gapped boundaries and boundary defects and list some open problems.
\end{abstract}

\maketitle


\section{Introduction}
\label{sec:intro}

A second quantum revolution in and around the construction of a large scale quantum computer is gaining momentum.  A unique approach is topological quantum computation (TQC) based on topological phases of matter (TPMs) \cite{Free98, Kitaev97, FKLW}.  TPMs  are new materials which exhibit many-body quantum entanglement, and their topological orders are patterns of long-range entanglement encoded algebraically by anyon models.  TQC is maturing at the forefront of the second quantum revolution as a killer application of TPMs.  The hardware of an anyonic quantum computer will be a TPM that harbors non-abelian objects.  The low energy effective theory of a TPM is a unitary topological quantum field theory (TQFT) and the anyons are modeled by simple objects in the associated anyon model (mathematically unitary modular category). 
TPMs have degenerate ground states that are robust against any local perturbations.  Those  degenerate ground states are
topological degrees of freedom, which can be used to construct qubits. Information that is encoded into topological degrees of freedom is automatically immune to local errors.  It follows that protection of information from local interactions with the environment is conferred at a physical level, with no active error correction needed.

While universal anyons by braiding alone such as the Fibonacci anyon is theoretically a possibility, accessible anyons with current technology all belong to the class that is weakly integral (WI).  If the property F conjecture \cite{Naidu09} holds, then they cannot be universal for quantum computation by braiding alone.  Since we are interested in using the non-abelian statistics for quantum computation, we will focus on gates obtained from non-abelian objects.

Recent studies in topological phases of matter revealed that certain topological phases of matter also support gapped boundaries \cite{Bravyi98,Beigi11,KitaevKong}. Furthermore, if the TPM supports multiple types of gapped boundaries, it will support {\it boundary defects} between different gapped boundary types (physical theories have been developed in \cite{ Levin13, Bark13a,Bark13b,Bark13c, Kapustin14,Cong16a,Cong17a}). Therefore, it is natural to ask if these cousins of anyons can be employed for quantum information processing.  This is indeed the case \cite{Cong16a,Cong17b}.

In the UMC model of a 2D doubled topological order $\mathcal{B}=\mathcal{Z}(\mathcal{C})$, a stable gapped boundary is modeled by a Lagrangian algebra $\mathcal{A}$ in $\mathcal{B}$ \cite{Kitaev09,KitaevKong}.  The Lagrangian algebra $\mathcal{A}$ consists of a collection of bulk bosonic anyons that can be condensed to vacuum at the boundary, and the corresponding gapped boundary is a condensate of those anyons which behaves as a non-abelian anyon of quantum dimension $d_{\mathcal{A}}$.  Lagrangian algebras in $\mathcal{B}=\mathcal{A}(\mathcal{C})$ are in one-to-one correspondence with indecomposable module categories $\mathcal{M}$  over $\mathcal{C}$, which can also be used to label gapped boundaries \cite{Davydov12}.

A route to creating, manipulating, and measuring topological degeneracy for gapped boundaries in $\mathfrak{D}(\Z_3)$ in bilayer fractional quantum Hall states coupled to superconductors has been proposed \cite{Bark16}. Other experimentally reasonable designs are proposed to realize boundary defects in abelian fractional quantum Hall states \cite{clarke2013, cheng2012, lindner2012, Bark16, GGGG}. 

The property $F$ conjecture suggests that braidings alone for gapped boundaries would not be sufficient for universality.  This leads us to consider more powerful computational primitives, such as the generalized topological charge measurement in \cite{Cong16a}.  We add our generalized topological charge measurement to braidings to obtain universal quantum computation from gapped boundaries in $\mathfrak{D}(\Z_3)$ \cite{Cong17b}.

\section{Summary}
\label{sec:summary}

\begin{figure}
\centering
\includegraphics[width = 0.4\textwidth]{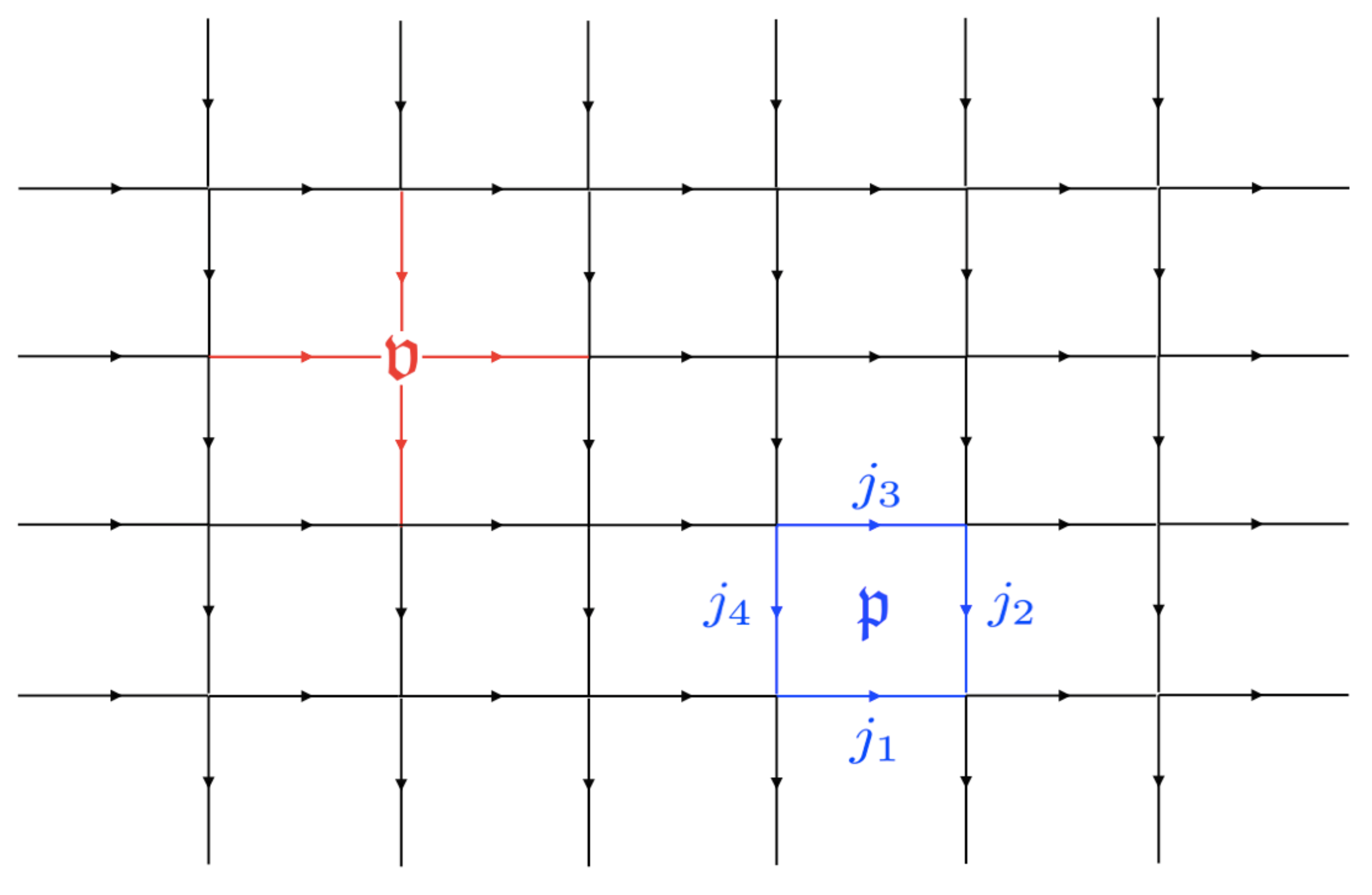}
\caption{Lattice for the Kitaev model. For simplicity of illustration and calculation, we use a square lattice, but in general, one can use an arbitrary lattice. If the group $G$ is non-abelian, it is necessary to define orientations on edges, as we have shown here. The edges $j$ and $j_1,...j_m$, used to obtain $A^g(v)$ and $B^h(p)$, are illustrated for this example of $v,p$.}
\label{fig:kitaev}
\end{figure}

Let $G$ be a finite group. Ref. \cite{Kitaev97} provides a local commuting projector Hamiltonian to realize the Dijkgraaf-Witten gauge theory $\mfD(G)$ from an arbitrary lattice of data qudits in a Hilbert space spanned by orthonormal basis $\{ g \in G \}$. In these theories, gapped boundaries are classified by a subgroup $K \subseteq G$ up to conjugation, and a 2-cocycle $\omega \in H^2 (K,\C^\times)$ \cite{KitaevKong,Cong16a,Cong16b}. In Refs. \cite{Cong16a,Cong16b}, we extend the Hamiltonians of \cite{Kitaev97,Bombin08,Beigi11}, to realize gapped boundaries in $\mfD(G)$ when the cocycle is trivial. The extension is as follows\footnote{The Hamiltonians published in Ref. \cite{Cong16a,Cong16b} contain redundancy and inaccuracy. The following is a corrected version.}:

Kitaev's famous toric code paper \cite{Kitaev97} presents a Hamiltonian to realize the Dijkgraaf-Witten theory based on any finite group $G$ on a general directed lattice and perform topological quantum computation. For simplicity of illustration and calculation, we will assume throughout our paper that the lattice is the square lattice in the plane; however, it is clear that all of the developed theory here extends to arbitrary lattices. In this model, a data qudit is placed on each edge of the lattice, as shown in Fig. \ref{fig:kitaev}. The Hilbert space for each qudit has an orthonormal basis given by $\{\ket{g}: g \in G\}$, so the total Hilbert space is $\mathcal{L}=\otimes_{e}\mathbb{C}[G]$.

As discussed in Ref. \cite{Kitaev97}, a Hamiltonian is used to transform the high-dimensional Hilbert space of all data qudits into a topological encoding. This Hamiltonian is built from several basic operators on a single data qudit:

\begin{equation}
\label{eq:L}
L^{g_0}_+ \ket{g} = \ket{g_0g}
\end{equation}
\begin{equation}
L^{g_0}_- \ket{g} = \ket{gg_0^{-1}}
\end{equation}
\begin{equation}
T^{h_0}_+ \ket{h} = \delta_{h_0,h}\ket{h}
\end{equation}
\begin{equation}
\label{eq:T}
T^{h_0}_- \ket{h} = \delta_{h_0^{-1},h}\ket{h}
\end{equation}

\noindent
where $\delta_{i,j}$ is the Kronecker delta function. These operators are defined for all elements $g_0,h_0 \in G$, and provide a faithful representation of the left/right multiplication and comultiplication in the Hopf algebra $\C[G]$. Using these operators, local gauge transformations and magnetic charge operators are defined as follows, on each vertex $v$ and plaquette $p$ \cite{Kitaev97}:

\begin{equation}
\label{eq:kitaev-vertex-g-term}
A^{g}(v,p) = A^{g}(v) = \prod_{j \in \text{star}(v)} L^g(j,v)
\end{equation}

\begin{equation}
\label{eq:kitaev-plaquette-h-term}
B^h(v,p) = \sum_{h_1 \cdots h_k = h} \prod_{m=1}^k T^{h_m}(j_m, p)
\end{equation}

Here, $j_1, ..., j_k$ are the boundary edges of the plaquette $p$ in counterclockwise order originating from the vertex $v$ (see Fig. \ref{fig:kitaev}), and $L^g$ and $T^h$ are defined as follows: if $v$ is the origin of the directed edge $j$, $L^g(j,v) = L^g_-(j)$, otherwise $L^g(j,v) = L^g_+(j)$; if $p$ is on the left (resp., right) of the directed edge $j$, $T^h(j,p) = T^h_-(j)$ ($T^h_+(j)$) \cite{Kitaev97}.

Note that since the $A^g(v)$ satisfy $A^g(v) A^{g'}(v) = A^{gg'}(v)$, the set of all $A^g(v)$ (for fixed $v$) form a representation of $G$ on the entire Hilbert space $\mathcal{L}=\otimes_{e}\mathbb{C}[G]$ of all data qudits \cite{Bombin08}.

Finally, two more linear combinations of these $A^g$ and $B^h$ operators are required to define the Hamiltonian:

\begin{equation}
\label{eq:kitaev-vertex-term}
A(v) = \frac{1}{|G|} \sum_{g \in G} A^g(v,p) 
\end{equation}

\begin{equation}
\label{eq:kitaev-plaquette-term}
B(p) = B^1(v,p).
\end{equation}

\noindent
The Hamiltonian\footnote{We call this Hamiltonian $H_{(G,1)}$, as this model is the Dijkgraaf-Witten theory with trivial cocycle (twist). In general, this Hamiltonian may be twisted by a 3-cocycle $\omega \in H^3(G,\C^\times)$, and may be written as $H_{(G,\omega)}$.} for the Kitaev model is then defined as

\begin{equation}
\label{eq:kitaev-hamiltonian}
H_{(G,1)} = \sum_v (1-A(v)) + \sum_p (1-B(p))
\end{equation}

\begin{figure}
\centering
\includegraphics[width = 0.4\textwidth]{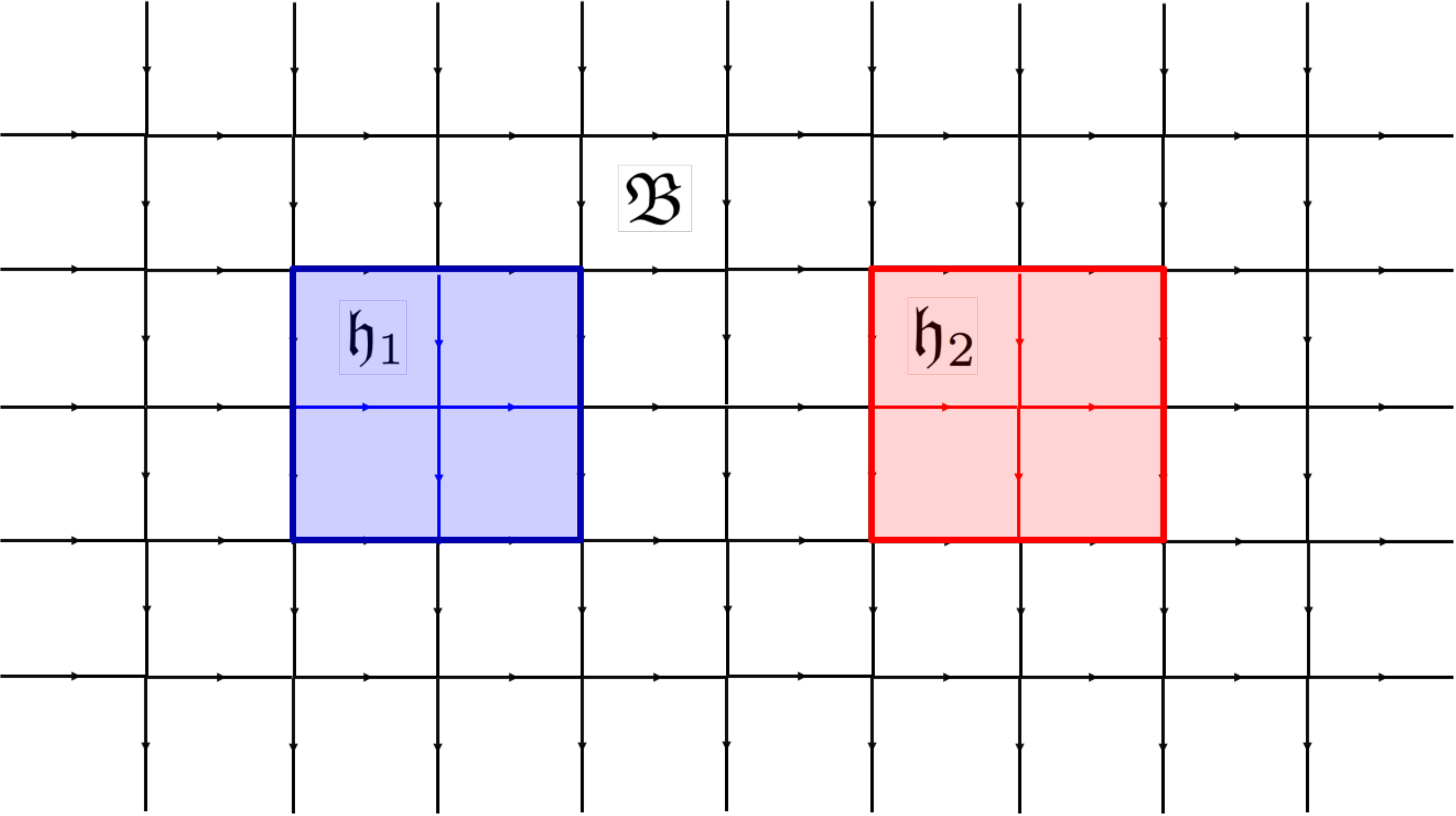}
\caption{Square lattice with boundaries, used to define the gapped boundary Hamiltonian. The Hamiltonian $H_{\text{G.B.}}$ is a sum of two terms, where the boundary Hamiltonian $H_{(G,1)}^{(K,1)}$ acts on edges in the hole, and the bulk Hamiltonian $H_{(G,1)}$ acts on vertices and plaquettes in the bulk.}
\label{fig:boundary}
\end{figure}

We now extend this Hamiltonian to surfaces with boundaries (e.g. in Fig. \ref{fig:boundary}). We will consider the model in which a gapped boundary is determined by a subgroup $K \subseteq G$ up to conjugation. In general, as shown in Ref. \cite{Beigi11}, a boundary is determined by both $K$ and a 2-cocycle $\phi \in H^2(K,\C^\times)$, and it is straightforward to generalize our results.

Let us now define some new projector terms, which act on the edges of the lattice:



\begin{equation}
\label{eq:LK}
L^K(e) := \frac{1}{|K|} \sum_{k \in K} (L^k_+(e) + L^k_- (e))
\end{equation}

\begin{equation}
\label{eq:TK}
T^K(e) := \sum_{k \in K} T^k_+(e)
\end{equation}

These projectors generalize the ones given in Ref. \cite{Bombin08} to potentially non-normal subgroups $K$. The definitions of $L^k$ and $T^k$ for Eqs. (\ref{eq:LK}-\ref{eq:TK}) are based on Eqs. (\ref{eq:L}-\ref{eq:T}). The choice of using only $T_+$ in Eq. (\ref{eq:TK}) is arbitrary, as using only $T_-$ would yield the same operator.

Following Ref. \cite{Bombin08}, we can now define the following Hamiltonian\footnote{As before, we write $H^{(K,1)}_{(G,1)}$ to leave room for the generalized version, where a boundary depends also on a 2-cocycle $\phi$ of $K$.}: 

\begin{equation}
\label{eq:bd-hamiltonian-K}
H^{(K,1)}_{(G,1)} = 
\sum_e ((1-T^K(e)) + (1-L^K(e))
\end{equation}

It is important to note that as in the Hamiltonian (\ref{eq:kitaev-hamiltonian}), all terms in this Hamiltonian commute with each other. Hence $H^{(K,1)}_{(G,1)}$ is also gapped. Using this, a commuting Hamiltonian was defined to realize $n$ gapped boundaries given by subgroups $K_1, ... K_n$ (see Fig. \ref{fig:boundary} for example):

\begin{equation}
\label{eq:gapped-bds-hamiltonian}
H_{\text{G.B.}} = H_G(\mathfrak{B}) + \sum_{i=1}^{n} H^{K_i}_{G}(\mathfrak{h}_i).
\end{equation}

In the bulk of the Dijkgraaf-Witten theory $\mfD(G)$ based on input group $G$, elementary excitations are classified by pairs $(C,\pi)$, where $C$ is a conjugacy class of $G$, and $\pi$ is an irreducible representation of the centralizer $E(C)$ of $C$. This can be shown directly from the Hamiltonian by introducing the notion of {\it ribbon operators} \cite{Kitaev97}. Refs. \cite{Cong16a,Cong16b} extend the definition of bulk ribbon operators to ribbon operators along the boundary, to show that the elementary excitations along a gapped boundary given by subgroup $K$ are classified by pairs $(T,R)$, where $T \in K\backslash G / K$ is a double coset, and $R$ is an irreducible representation of the stabilizer $K^{r_T} = K \cap r_T K r_T^{-1}$ ($r_T \in T$ is any representative of the double coset). In particular, a correspondence is given between the bulk and boundary ribbon operators which characterizes the physical bulk-to-boundary condensation procedure \cite{Cong16a,Cong16b}. The correspondence describes how a bulk anyon $(C,\pi)$ becomes a direct sum of boundary excitations $(T,R)$ upon condensation to the boundary.

\begin{figure}
\centering
\includegraphics[width = 0.4\textwidth]{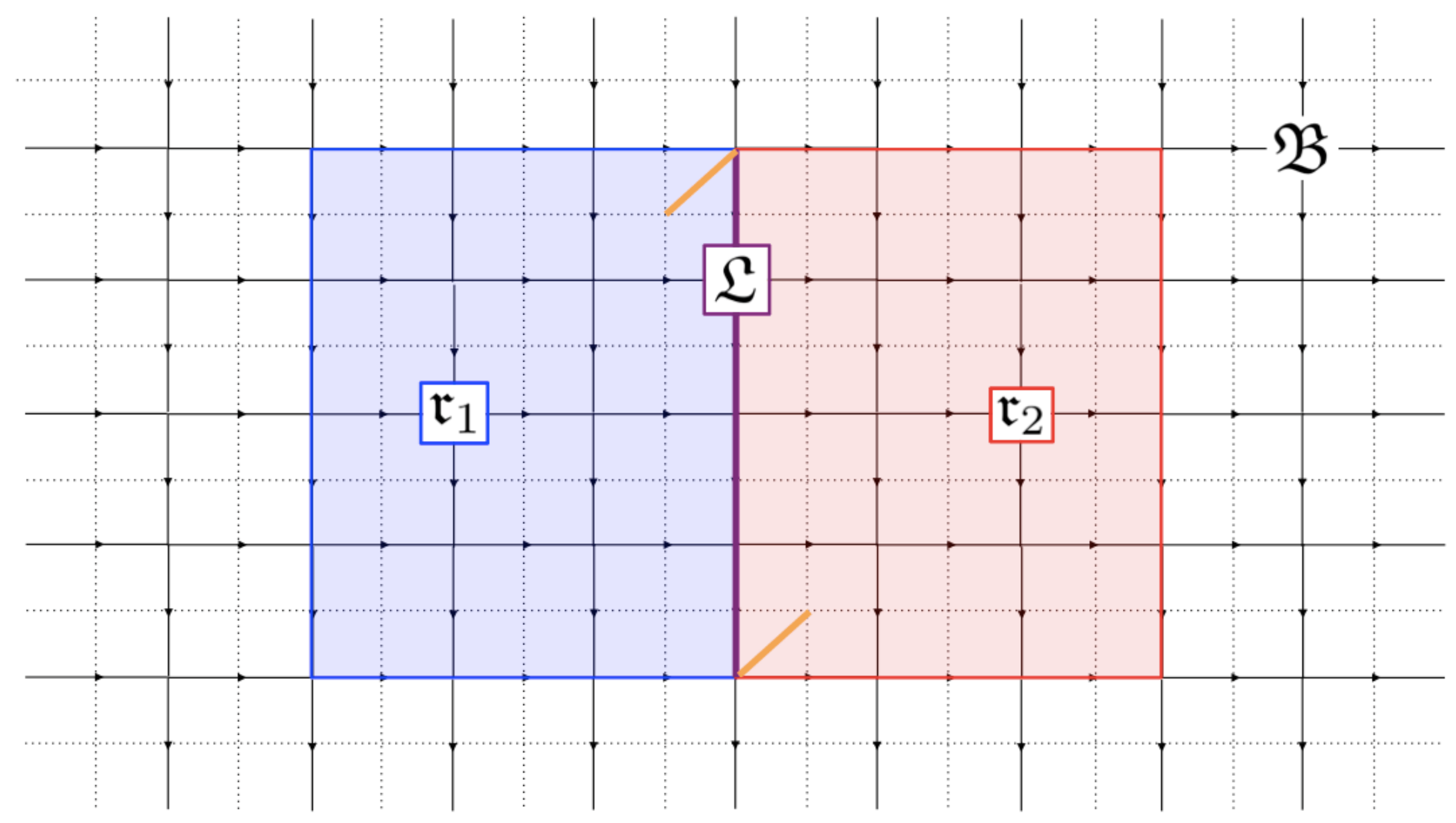}
\caption{Boundary defects. The boundary defects are located at the junction of different boundary types.}
\label{fig:defects}
\end{figure}

More generally, Refs. \cite{Cong16a,Cong17a} also define a local commuting projector Hamiltonian to realize boundary defects, which occur at the junction points of two different boundary types (see Fig. \ref{fig:defects}). In the Dijkgraaf-Witten theory based on a finite group $G$, the defects between gapped boundaries given by $K_1$ and $K_2$ (and trivial cocycles) are classified by pairs $(T,R)$, where $T \in K_1 \backslash G / K_2$ is a double coset, and $R \in (K_1, K_2)^{r_T}_{\text{ir}}$ is an irreducible representation of the stabilizer $(K_1, K_2)^{r_T} = K_1 \cap r_T K_2 r_T^{-1}$.

While these local commuting projector Hamiltonians provide a good way to understand gapped boundaries physically, a more elegant and abstract way of describing them is through the language of category theory. In particular, consider a Turaev-Viro TQFT which is a Drinfeld center $\B = \mZ(\mC)$, for some unitary fusion category $\mC$. The gapped boundaries in this theory are given by indecomposable module categories $\M$ of $\mC$, or equivalently, Lagrangian algebras $\A$ in $\B$. Excitations on such a gapped boundary are given by the fusion category $\Fun_\mC(\M,\M)$ \cite{KitaevKong}.

Similarly, while ribbon operators provide one way to describe bulk-to-boundary condensation, Refs. \cite{Cong16a,Cong16b} also present this condensation procedure as a quotient functor followed by an idempotent completion as in \cite{Muger03} (overall, a tensor functor):

\begin{equation}
\label{eq:condensation-quotient-IC}
F: \mZ(\mC) = \B \xrightarrow{\text{quotient}} \B/\A = \widetilde{\mQ} \xrightarrow{\text{I.C.}}  {\mQ} = \Fun_\mC(\M,\M).
\end{equation}

Let $\{M_i\}_{i=1}^n$ be a complete collection of gapped boundary types in the theory, corresponding to Lagrangian algebras $\{ A_i \}_{i=1}^n$. Then one can form an $n \times n$ multi-fusion category which describes all possible boundary excitations and defects between gapped boundaries:

\begin{equation}
\mathfrak{C} = \{ \mC_{ij} := \Fun_\mC (\M_i, \M_j) : i,j = 1,2, ... n\}.
\end{equation}

This multi-fusion category can be used to compute important topological properties of boundary excitations and defects, such as their quantum dimension and fusion rules.

As discussed in Ref. \cite{Cong17a}, the bulk-to-boundary condensation procedure of Eq. (\ref{eq:condensation-quotient-IC}) can also be generalized to boundary defects. In this case, the bulk category is not simply $\B = \mZ(\mC)$, and some further setup is required. First, a global symmetry group of the category $\B$ is a group $G$ with a homomorphism $\rho: G \rightarrow \text{Aut}^{\text{br}}_{\otimes}(\B)$ from $G$ to the braided tensor auto-equivalences of $\B$. Under $\rho$, each element of $g$ is sent to a permutation $\rho_g$ of the simple objects of $\B$. In particular, $\rho_g$ can also act on a Lagrangian algebra (i.e. gapped boundary): $\rho_g(\A_i) = \A_{j_g}$.

Let $G$ be a global symmetry group of our anyon theory $\B = \mZ(\mC)$. Then $\B$ can be extended to a $G$-graded fusion category $\B_G = \oplus_{g \in G} \B_g$ which describe all bulk anyons and bulk symmetry defects corresponding to the $G$-symmetry \cite{Barkeshli14}. The generalized bulk-to-boundary condensation procedure, called {\it crossed condensation} \cite{Cong17a}, is then

\begin{equation}
\label{eq:crossed-condensation-quotient-IC-G}
F: \B_{G} \xrightarrow{\text{quotient}} \B_{G}/\A_i = \widetilde{\mQ}(G, \A_i)
\xrightarrow{\text{I.C.}}  {\mQ}(G, \A_i) = \oplus_{g \in G} \Fun_{\mC}(\M_i, \M_{j_g}).
\end{equation}

\noindent
Here, $\M_{j_g}$ is the indecomposable module category corresponding to the Lagrangian algebra $\A_{j_g} = \rho_g(\A_i)$, and as before, I.C. denotes the idempotent completion. Hence, bulk anyons condense to excitations on the boundary $\A_i$, while bulk symmetry defects in the flux sector $\B_g$ condense to boundary defects in the category $\mC_{i j_g}$.

Using this understanding of crossed condensation, Ref. \cite{Cong17a} provides an understanding for the projective braiding of boundary defects in ${\mQ}(G, \A_i)$. Specifically, the tensor functor $F$ corresponding to crossed condensation has an adjoint $I$, which sends boundary defects to their bulk symmetry defect counterparts. Because these bulk symmetry defects have a well-defined $G$-crossed braiding, this also gives a projective ($G$-crossed) braiding for boundary defects. As specific cases, Ref. \cite{Cong17a} shows how this understanding explains the projective braid statistics of Majorana and parafermion zero modes when realized as boundary defects in topological phases \cite{lindner2012}, which is an important application to topological quantum computation.

\begin{figure}
\centering
\includegraphics[width = 0.08\textwidth]{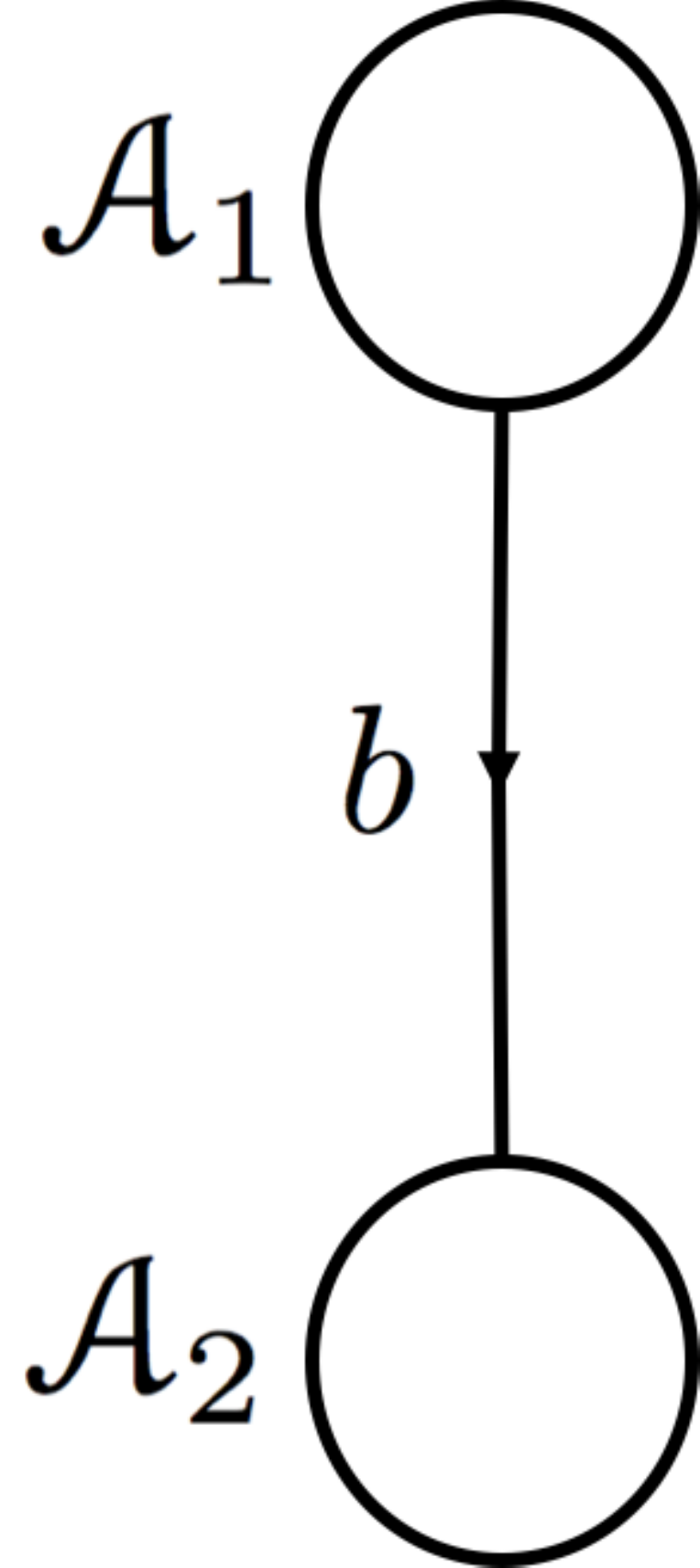}
\caption{Qudit encoding. The ground state is given by the anyons that can tunnel from one boundary to another, without any energy cost.}
\label{fig:encoding}
\end{figure}

Finally, Refs. \cite{Cong16a,Cong17b} provide qudit encodings based on gapped boundaries. In particular, given two gapped boundaries (Lagrangian algebras) $\A_i$ and $\A_j$ in the plane with total charge vacuum, the ground state is given by the Hilbert space

\begin{equation}
\Hom(\A_i, \A_j) \cong \C^d
\end{equation}

\noindent
for some $d$. This Hilbert space may be used to encode a $d$-dimensional qudit, as shown in Fig. \ref{fig:encoding}.

\floatsetup[figure]{style=plain,subcapbesideposition=bottom}
\begin{figure}
     \centering
        \sidesubfloat[]{%
            \includegraphics[width=0.25\textwidth]{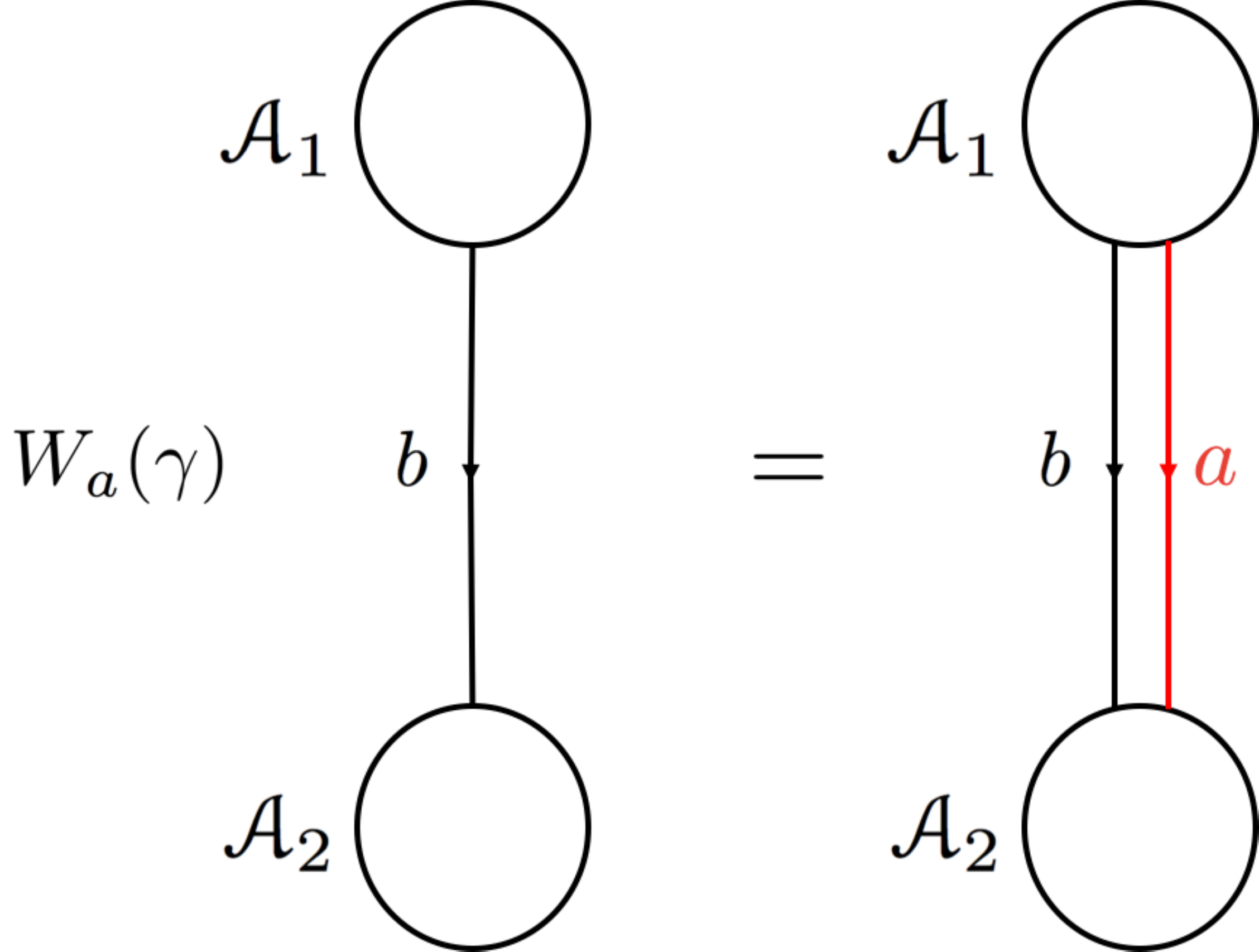}\label{fig:tunnel}
        }
        \sidesubfloat[]{%
           \includegraphics[width=0.25\textwidth]{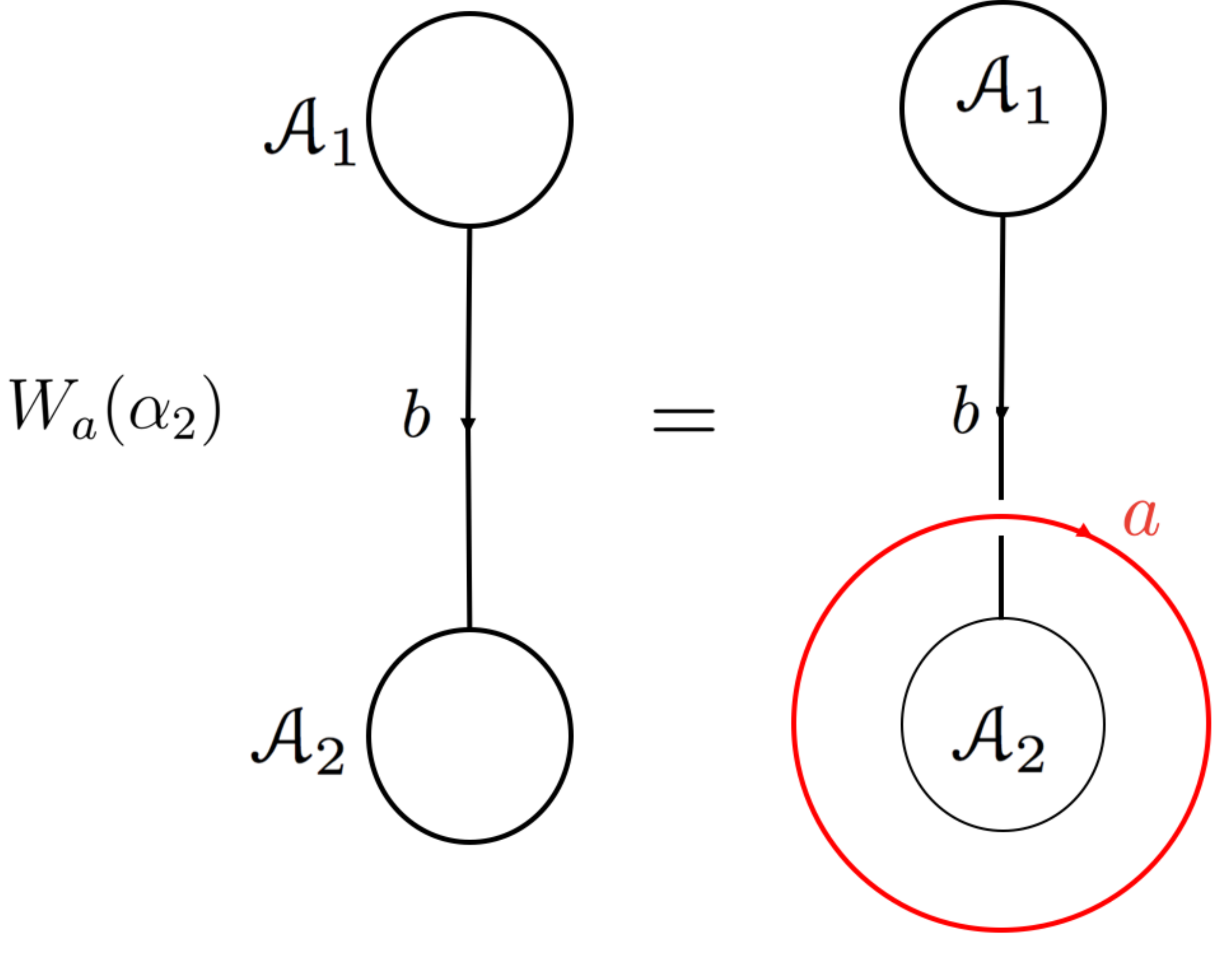}\label{fig:loop}
        }
		\sidesubfloat[]{%
           \includegraphics[width=0.2\textwidth]{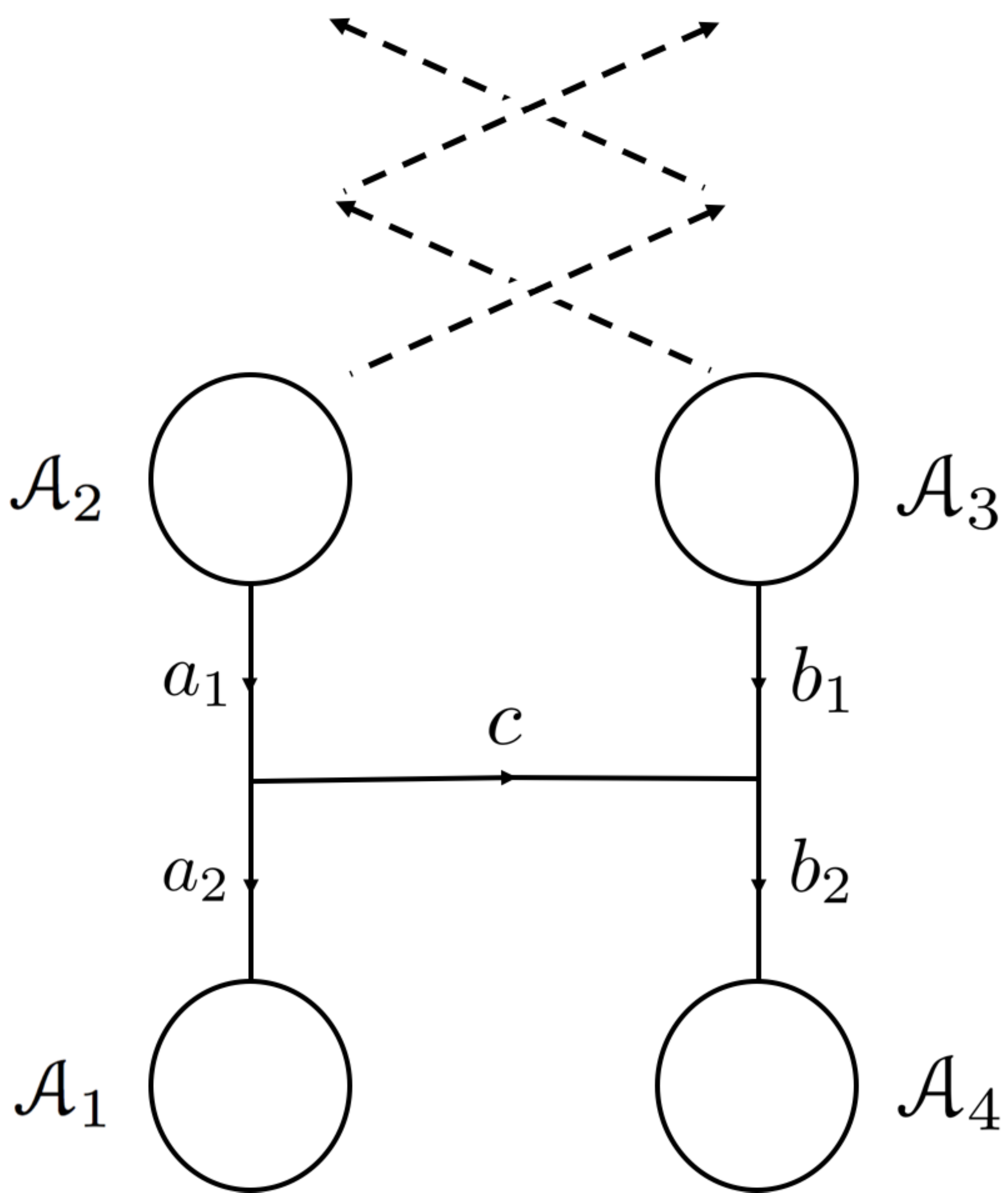}\label{fig:braid}
        }
    \caption{(A) Tunnel-$a$ operator. (B) Loop-$a$ operator. (C) Braiding gapped boundaries.}%
\end{figure}

Given such topological qudits, there are several possible operations one can perform on the associated Hilbert space. First, one can tunnel an anyon $a$ from the boundary $\A_i$ to the boundary $\A_j$, if the antiparticle $\overbar{a}$ condenses on $\A_i$, and $a$ condenses on $\A_j$. This is illustrated in Fig. \ref{fig:tunnel}. Second, one can loop any bulk anyon around one of the gapped boundaries, as shown in Fig. \ref{fig:loop}. Third, one can braid gapped boundaries around each other (Fig. \ref{fig:braid}). 

The final operation that can be performed is a new primitive introduced in Refs. \cite{Cong16a,Cong17b} called {\it topological charge measurement}, which generalizes the topological charge projection discussed in \cite{Barkeshli16}. This is a very general primitive, which measures arbitrary Hermitian operators along Wilson lines and loops; however, in the cases of interest, the required measurement is usually much more physical. For instance, Ref. \cite{Cong16b} presents a universal quantum gate set, where the required topological charge measurement can be implemented in a symmetry-protected fashion.

Using these four topological operations, Refs. \cite{Cong16a,Cong17b} show how universal quantum computation can be achieved using purely the gapped boundaries of the quantum double model $\mfD(\Z_3)$. This is a very notable illustration of the additional computation power of gapped boundaries over anyons, as $\mfD(\Z_3)$ is a purely abelian theory where anyon braidings are trivial. 

\section{Open questions}
\label{sec:open-questions}

Gapped boundaries and boundary defects are non-abelian objects beyond anyons \cite{Wang17}.  Most questions about non-abelian anyons have analogues for gapped boundaries and boundary defects plus some new ones.  In this section, we compile some questions.

\subsection{Beyond finite groups}

In Refs. \cite{Cong16a,Cong16b,Cong17a}, we provided a Hamiltonian realization for gapped boundaries and boundary defects in $(2+1)$D Dijkgraaf-Witten theories for any finite group $G$, based on Kitaev's quantum double models \cite{Kitaev97}. 

The first problem is to generalize our theory to include nontrivial cocycles such as $(K, \omega)$ for the boundary of twisted DW theories.  More generally, the Levin-Wen Hamiltonian \cite{Levin04} can realize a much larger class of $(2+1)$-D TQFTs. In this model, the representation category of $G$ is generalized to any unitary fusion category $\mathcal{C}$. By \cite{Chang14}, Kitaev's model can be made to realize the same TQFTs as the Levin-Wen model when $G$ is replaced by a {\it weak Hopf algebra}. An interesting direction is to generalize the gapped boundary and boundary defect Hamiltonians of Refs. \cite{Cong16a,Cong16b,Cong17a} to the Levin-Wen model by replacing the input group with a weak Hopf algebra.

\subsection{Beyond two dimensions}

Another interesting direction is to consider higher-dimensional analogs of topological phases and their gapped boundaries. One specific case is to examine the Walker-Wang model for the bulk of $(3+1)$-D TQFTs \cite{WW}, and study this model in the presence of gapped boundaries and boundary defects.  Even more generally, one could extend the models to the TQFTs and lattice model based on $G$-crossed braid categories \cite{Cui, WW16}.  

\subsection{Beyond symmetry}

There are general defects beyond those associated with topological symmetries.  A notable example is the $A+C+D/A+B+2F$ boundary defect in the $\mfD(S_3)$ theory, as discussed in Ref. \cite{Cong17a}. An extension to such defects is an interesting direction.

\subsection{Stability of gapped boundaries}

The most interesting question that we have not touched on is the stability of the topological degeneracy in our model.  Once our Hamiltonian moves off the fixed-point, finite-size splitting of the degeneracy would occur.  It would be interesting to study the energy splitting of the ground state degeneracies of the Hamiltonians $H_{\text{G.B.}}$ and $H_{\text{dft}}$ of Refs. \cite{Cong16a,Cong16b,Cong17a} numerically under small perturbations.

\subsection{Domain walls}

Many physics papers have studied gapped domain walls between different topological phases. While gapped boundaries are often considered as a special case of gapped domain walls, by the folding trick \cite{KitaevKong}, they also completely cover the domain wall theory mathematically. Physically, however, it is still interesting to analyze the general gapped domain walls following our work.

\subsection{Applications to TQC}

There are several open questions in our understanding of topological quantum computation with gapped boundaries and boundary defects.

\begin{enumerate}
\item First, one important open question is to obtain a systematic method to perform TQC with boundary defects, including an efficient qudit encoding scheme and list of topologically protected operations. Furthermore, one should obtain the physical realizations of such quantum gates.
\item Ref. \cite{Cong17a} uses fusion channels to understand the degeneracy of many boundary defects on a ring. It would be useful to develop ribbon operator techniques to understand this degeneracy, as such a picture would give rise to a clearer picture of topological operations that can be performed on boundary defects. 
\item An interesting question is to determine computational power for a simple theory such as the $\Z_2$ toric code with gapped boundaries, symmetry defects, boundary defects all included.
\item Ref. \cite{Escobar17} analyzes the braid group representations resulting from braiding gapped boundaries. An interesting and more general question is to examine the gapped boundaries (and potentially also the boundary defects) of the Dijkgraaf-Witten theory $\mfD(S_3)$ with entire braid group representation, tunneling operators (i.e. some non-unitary operators), and determine whether this set of operations is universal.
\item Finally, another important implementation scheme for TQC with gapped boundaries is through quantum circuits and the surface code (e.g. Refs. \cite{Fowler12,Cong16a}). Ref. \cite{Fowler12} discusses only the special case where the bulk is the $\Z_2$ toric code, and Ref. \cite{Cong16a} provides an outline in Chapter 4. It would be interesting to find a more detailed implementation for general gapped boundaries.
\end{enumerate}

\subsection{Implementing universal gate set}

Ref. \cite{Cong17a} presents a symmetry-protected implementation of the topological charge measurement required for universality. One very interesting problem would be to obtain a purely topological implementation, as this would give a purely topological gate set using only an abelian theory.


\vspace{4mm}
\begin{thebibliography}{99}
\bibliographystyle{alpha}
\bibitem{Barkeshli14}
M. Barkeshli, P. Bonderson, M. Cheng, Z. Wang. {\it Symmetry, defects, and gauging of topological phases}. arXiv preprint arXiv:1410.4540 (2014).
\bibitem{Bark16}M. Barkeshli. {\it Charge $2 e/3$ superconductivity and topological degeneracies without localized zero modes in bilayer fractional quantum Hall states}. Phys. Rev. Lett. {\bf 117}, 096803 (2016). 
\bibitem{Barkeshli16}
M. Barkeshli, M. Freedman. {\it Modular transformations through sequences of topological charge projections}. arXiv preprint arXiv:1602.01093 (2016).
\bibitem{Bark13a}M. Barkeshli, C.-M. Jian, X.-L. Qi. {\it Twist defects and projective non-Abelian braiding statistics}. Physical Review B {\bf 87}(4), 045130 (2013).
\bibitem{Bark13b}M. Barkeshli, C.-M. Jian, X.-L. Qi. {\it Theory of defects in Abelian topological states}. Physical Review B {\bf 88}(23), 235103 (2013).
\bibitem{Bark13c}M. Barkeshli, C.-M. Jian,  X.-L. Qi. {\it Classification of topological defects in Abelian topological states}. Physical Review B {\bf 88}(24), 241103 (2013).
\bibitem{Beigi11}
S. Beigi, P. W. Shor, D. Whalen. {\it The quantum double model with boundary: condensations and symmetries}. Communications in Mathematical Physics {\bf 306}(3), 663-694 (2011).
\bibitem{Bombin08}
H. Bombin, M. A. Martin-Delgado. {\it Family of non-Abelian Kitaev models on a lattice: Topological condensation and confinement}. Phys. Rev. B {\bf 78}, 115421 (2008).
\bibitem{Bravyi98} 
S. B. Bravyi, A. Y. Kitaev. {\it Quantum codes on a lattice with boundary}. arXiv:quant-ph/9811052 (1998).
\bibitem{Chang14}L. Chang. {\it Kitaev models based on unitary quantum groupoids}. J. Math. Phys. 55.4 (2014).
\bibitem{cheng2012}
M. Cheng. Phys. Rev. B, {\it Superconducting Proximity Effect on the Edge of Fractional Topological Insulators}. {\bf 86}, 195126 (2012)
\bibitem{clarke2013}  
D.~J. Clarke, J. Alicea and K. Shtengel.  {\it Exotic non-Abelian anyons from conventional fractional quantum Hall states}. {Nature Comm.} {\bf 4}, 1348 (2013).
\bibitem{Cong16a}
I. Cong, M. Cheng, Z. Wang. {\it Topological quantum computation with gapped boundaries.} arXiv preprint arXiv:1609.02037 (2016).
\bibitem{Cong16b}
I. Cong, M. Cheng, Z. Wang. {\it Hamiltonian and Algebraic Theories of Gapped Boundaries in Topological Phases of Matter}. Commun. Math. Phys. (2017). doi:10.1007/s00220-017-2960-4
\bibitem{Cong17a}
I. Cong, M. Cheng, Z. Wang. {\it On Defects Between Gapped Boundaries in Two-Dimensional Topological Phases of Matter}. arXiv preprint arXiv:1703.03564 (2017).
\bibitem{Cong17b}
I. Cong, M. Cheng, Z. Wang. {\it Universal Quantum Computation with Gapped Boundaries}. arXiv preprint arXiv:1707.05490 (2017). To appear in {\it Phys. Rev. Lett.}
\bibitem{Cui}S.-X.  Cui. {\it Higher Categories and Topological Quantum Field Theories.} arXiv preprint arXiv:1610.07628 (2016).
\bibitem{Davydov12}
A. Davydov, M. M{\"u}ger, D. Nikshych, V. Ostrik. {\it The Witt group of non-degenerate braided fusion categories.} Journal f{\"u}r die reine und angewandte Mathematik (2012). doi: 10.1515/crelle.2012.014.
\bibitem{Escobar17}
N. Escobar-Vel{\'a}squez, C. Galindo, Z. Wang. {\it Braid Group Representations from Braiding Gapped Boundaries of Dijkgraaf-Witten Theories}. arXiv preprint arXiv:1707.03884 (2017).
\bibitem{Fowler12}
A.G. Fowler, M. Mariantoni, J. M. Martinis, A. N. Cleland. {\it Surface codes: Towards practical large-scale quantum computation}. Phys. Rev. A {\bf 86}(3), 032324 (2012).
\bibitem{Free98}M.H. Freedman. {\it P/NP, and the quantum field computer}. Proceedings of the National Academy of Sciences {\bf 95}(1), 98-101 (1998).
\bibitem{FKLW}M. Freedman, A. Kitaev, M. Larsen, Z. Wang. {\it Topological quantum computation}. Bulletin of the American Mathematical Society, {\bf 40}(1), 31-38 (2003).
\bibitem{GGGG}S. Ganeshan, A. V. Gorshkov, V. Gurarie, V. M. Galitski. {\it Exactly soluble model of boundary degeneracy}. arXiv preprint arXiv:1604.02089.
\bibitem{Kapustin14}
A. Kapustin, {\it Ground-state degeneracy for Abelian anyons in the presence of gapped boundaries}, Phys. Rev. B {\bf 89}, 125307 (2014)
\bibitem{KitaevKong}
A. Kitaev, L. Kong, {\it Models for Gapped Boundaries and Domain Walls}. Commun. Math. Phys. {\bf 313}, 351-373 (2012). doi: 10.1007/s00220-012-1500-5.
\bibitem{Kitaev97}
A. Y. Kitaev. {\it Fault-tolerant quantum computation by anyons}. Ann. Phys. {\bf 303}(2) (2003).
\bibitem{Kitaev09}A. Kitaev. Bose-condensation and edges of topological quantum phases.  Talk at modular categories and applications, Indiana University, March 19-22, 2009.
\bibitem{Levin04}
M. A. Levin, X.-G. Wen. {\it String-net condensation: A physical mechanism for topological
phases}. Phys. Rev. B {\bf 71}, 045110 (2005).
\bibitem{Levin13}
M. Levin, {\it Protected edge modes without symmetry}. Phys. Rev. X {\bf 3}, 021009 (2013).
\bibitem{lindner2012}
N.~H. Lindner, E. Berg, G. Refael and A. Stern. {\it Fractionalizing Majorana fermions: non-abelian statistics on the edges of abelian quantum Hall states}. Phys. Rev. X, {\bf 2}, 041002 (2012).
\bibitem{Muger03}
M. M{\"u}ger. {\it Galois extensions of braided tensor categories and braided crossed G-categories}. Journal of Algebra {\bf 277} 256–281 (2004).
\bibitem{Naidu09}
D. Naidu, E. C. Rowell. {\it A finiteness property for braided fusion categories.} Algebr. Represent. Theor. {\bf 14}, 837 (2011). doi:10.1007/s10468-010-9219-5
\bibitem{WW}
K. Walker, Z. Wang. {\it $(3+1)$-TQFTs and Topological Insulators.} Frontiers of Physics, {\bf 7}(2), 150-159 (2012).
\bibitem{WW16}D.-J. Williamson, and Z. Wang. {\it Hamiltonian models for topological phases of matter in three spatial dimensions.} Annals of Physics 377 (2017): 311-344.
\bibitem{Wang17}Z. Wang, {\it Beyond anyons}, arXiv:1710.00464.
\end{thebibliography}
\end{document}